\documentclass[useAMS,usenatbib]{mn2e}
\usepackage{txfonts}
\usepackage{natbib}
\usepackage[all]{xy}
\usepackage[dvips]{graphicx}

\usepackage{hyperref}
\usepackage[usenames,dvipsnames]{color}
\definecolor{menublue}{rgb}{0.0,0.0,0.5}
\definecolor{citegreen}{rgb}{0.0,1.0,0.0}
\definecolor{urlred}{rgb}{1.0,0.0,0.0}
\hypersetup{bookmarksopen,pdfstartview={FitH},colorlinks=true, breaklinks=true,menucolor=menublue,urlcolor=urlred,citecolor=citegreen,linkcolor=blue}
\usepackage[all]{hypcap}

\def\del#1{{}}

\newcommand{\ltsima}{$\; \buildrel < \over \sim \;$}
\newcommand{\lsim}{\lower.5ex\hbox{\ltsima}}
\newcommand{\gtsima}{$\; \buildrel > \over \sim \;$}
\newcommand{\gsim}{\lower.5ex\hbox{\gtsima}}
\newcommand{\bra}{\langle}
\newcommand{\ket}{\rangle}

\newcommand{\dd}{\mathrm{d}}

\newcommand{\trace}{\mathrm{tr}}

\newcommand{\dirac}{\delta_D}
\newcommand{\likelihood}{\mathcal{L}}

\title[BAOs with lensing]
{Detecting baryon acoustic oscillations by 3d weak lensing}
\author[A. Grassi and B. M. Sch{\"a}fer]{
Alessandra Grassi\thanks{e-mail: sandri@ari.uni-heidelberg.de} and Bj{\"o}rn Malte Sch{\"a}fer\\
Zentrum f{\"ur} Astronomie der Universit{\"a}t Heidelberg, Philosophenweg 12, 69120 Heidelberg, Germany }

\begin{document}
\pagerange{\pageref{firstpage}--\pageref{lastpage}}
\pubyear{2013}
\maketitle
\label{firstpage}

\begin{abstract}
We investigate the possibility of detecting baryon acoustic oscillation features in the cosmic matter distribution by 3d weak lensing. Baryon oscillations are inaccessible even to weak lensing tomography because of wide line-of-sight weighting functions and require a specialized approach via 3d shear estimates. We quantify the uncertainty of estimating the matter spectrum amplitude at the baryon oscillations wave vectors by a Fisher-matrix approach with a fixed cosmology and show in this way that future weak lensing surveys such as EUCLID and DES are able to pick up the first four wiggles, with EUCLID giving a better precision in the measurement. We also provide a detailed investigation of the correlation existing between errors and  of their scaling behavior with respect to survey parameters such as median redshift, error on redshift, error on the galaxy shape measurement, sky coverage, and finally with respect to the number of wiggles one is trying to determine.
\end{abstract}


\section{Introduction}
Baryon acoustic oscillation (BAO) features are modulations in the cosmic matter distribution on very large spatial scales of roughly $\sim100~\mathrm{Mpc}/h$ \citep[for a review, see][]{2010deot.book..246B}. These BAOs are the imprint of oscillations of the photon-baryon fluid in the early universe on the matter density field driven by gravity and the equation of state providing a restoring force, and they are observable in two primary channels: through the observation of anisotropies in the cosmic microwave background (CMB) and through galaxy surveys. The most important features such as their spatial scales, their signature in the CMB, their statistical properties, and their dependence on cosmological parameters is very well understood analytically \citep{1996ApJ...469..437S, 1996ApJ...471..542H, 2011PhRvD..84b3522M, 2012MNRAS.426.1280S}.
 
Concerning the determination of cosmological parameters, it is a fortunate situation that they are observable at high redshifts through the primary CMB and at much lower redshifts in the galaxy distribution. Due to the fact that BAOs provide a standard yardstick at two different cosmological epochs, it is possible to constrain the density parameters of cosmic fluids and the possible time evolution of their equation of state parameters in a geometric way, breaking degeneracies that may arise if the equations of state of cosmological fluids are allowed to change with time.

CMB observations carried out by the Cosmic Background Explorer \citep[COBE,][]{1994ApJ...436..423B, 1996ApJ...464L..21W} first revealed anisotropies in the CMB, but only the Wilkinson Microwave Anisotropy Probe \citep[WMAP,][]{2003ApJS..148..135H, 2007ApJS..170..288H, 2009ApJS..180..296N, 2011ApJS..192...16L} had sufficient angular resolution such that the BAO scale of $\sim100~\mathrm{Mpc}/h$ could be resolved at a comoving distance of $\sim10~\mathrm{Gpc}/h$, revealing temperature modulations of the CMB of the order $\Delta T/T_\mathrm{CMB}\simeq 10^{-5}$ at an angular scale of roughly $2^\circ$, with subsequent higher harmonics. Likewise, galaxy surveys have now reached sufficient depth and solid angle that BAO features could be detected as modulations of the galaxy density of the order 10\% in both radial and transverse directions. With the assumption of a galaxy biasing model, the longest wavelength BAO modes survive nonlinear structure formation to the present epoch \citep{1999MNRAS.304..851M} and will be targeted by future surveys for the precision determination of cosmological parameters \citep{2006MNRAS.366..884D, 2008MNRAS.383..755A, 2012ApJ...746..172L}, in particular dark energy \citep{2003ApJ...598..720S, 2007ApJ...664..675E}. Both avenues have contributed significantly to the estimation of cosmological parameters and to the selection of most plausible cosmological models.

Specifically, there are quite a number of detection reports with ongoing surveys, for instance with the Sloan Digital Sky Survey \citep[SDSS,][]{2005ApJ...633..560E, 2007MNRAS.378..852P, 2007ApJ...657...51P, 2010MNRAS.401.2148P, 2010ApJ...710.1444K, 2012arXiv1202.0090P,2012arXiv1202.0092M}, the 2-Degree Field Galaxy Redshift Survey \citep[2dFGRS,][]{2007MNRAS.381.1053P, 2011MNRAS.416.3017B}, the WiggleZ-survey \citep{wigglez} and Lyman-$\alpha$ data \citep{2012arXiv1211.2616B} with subsequent determination of cosmological parameters which confirm spatial flatness and the low matter density found by CMB observations, if flatness is assumed prior to the analysis. Recent studies \citep{2012arXiv1211.3976P, 2012arXiv1211.3741Z} were able to constrain neutrino masses. BAO modulations have been found as longitudinal as well as transverse modes in the galaxy density \citep{2009MNRAS.399..801G, 2010ApJ...719.1032K, 2009MNRAS.399.1663G} and their issues of model selection and parameter estimation have been addressed thoroughly \citep{2011MNRAS.412L..98C}.

The motivation for this paper is the fact that the detection of BAOs as a modulation feature in the galaxy field depends on the assumption of a biasing mechanism \citep{2009MNRAS.399..801G, 2010PhRvD..82j3529D} which relates the galaxy number density to the ambient density of dark matter is well as a control of redshift space distortions effects \citep{2007PASJ...59.1049N, 2009PhRvD..80l3503T} and it would be desirable to measure the dark matter density directly. Weak lensing would be a prime candidate for such a measurement, but the wide line-of-sight weighting functions cause the weak lensing signal to depend rather on the integral of the spectrum of cold dark matter (CDM) than on individual, localized features, even in the case of tomographic lensing surveys \citep{1999ApJ...522L..21H}. This is the reason why  investigate the sensitivity of 3d weak lensing \citep[3dWL,][]{2003MNRAS.343.1327H} for constraining the dark matter spectrum on BAO scales: 3dWL provides a direct estimate of the 3-dimensional matter distribution and gives Gaussian errors on the amplitude of the CDM spectrum in wavelength bands from sparsely sampled data \citep{2012A&A...539A..85L}. In this way, we aim to quantify the statistical precision at which 3dWL constraints the CDM spectrum at the BAO wavelengths, and the statistical significance for inferring the presence of one or more wiggles from 3dWL data relative to the null-hypothesis of absent wiggles.

After a short compilation of basic results concerning distances, structure growth, structure statistics, and conventional weak lensing in Sect.~\ref{sect_cosmology}, we recapitulate the main results of 3dWL in Sect.~\ref{sect_3dwl} and motivate its usage in constraining BAO wiggles. Our statistical approach and the estimation of statistical errors on the BAO measurement is given in Sect.~\ref{sect_wiggles}, followed by a discussion of our main results in Sect.~\ref{sect_summary}.

The reference cosmological model used is a spatially flat $w$CDM cosmology with Gaussian adiabatic initial perturbations in the matter distribution. The specific parameter choices are $\Omega_m = 0.25$, $n_s = 1$, $\sigma_8 = 0.8$ and $H_0=100\: h\:\mathrm{km}/\mathrm{s}/\mathrm{Mpc}$, with $h=0.72$. The dark energy equation of state is set to $w=-0.9$ and we assume the dark energy to be smooth. The baryon density $\Omega_b=0.04$ is used for correcting the CDM shape parameter and for predicting BAO-wiggle amplitudes and wave-vectors.

\section{cosmology and structure formation}\label{sect_cosmology}

\subsection{Dark energy cosmologies}
In spatially flat dark energy cosmologies with the present matter density $\Omega_m$, the Hubble function $aH(a)=\dd a/\dd t$ is given by
\begin{equation}
\frac{H^2(a)}{H_0^2} = \frac{\Omega_m}{a^{3}} + (1-\Omega_m)\exp\left(3\int_a^1\dd\ln a\:(1+w(a))\right),
\end{equation}
with the dark energy equation of state $w(a)$. A constant value $w\equiv -1$ corresponds to the cosmological constant. The relation between comoving distance $\chi$ and scale factor $a$ is given by
\begin{equation}
\chi = c\int_a^1\:\frac{\dd a}{a^2 H(a)},
\end{equation}
with the Hubble distance $\chi_H=c/H_0$ as the cosmological distance scale. Redshift $z$ and comoving distance are related by $\dd z/\dd\chi = H(z)/c$ .

\subsection{CDM power spectrum}
The linear CDM-density power spectrum $P(k)$ describes  the fluctuation amplitude of the Gaussian homogeneous density field $\delta$,
\begin{equation}
\bra\delta(\bmath{k})\delta(\bmath{k}^\prime)^*\ket=(2\pi)^3\dirac(\bmath{k}-\bmath{k}^\prime)P(k)
\propto k^{n_s}T^2(k),
\end{equation}
with the the spectral index $n_s$ and the transfer function $T(k)$. The restoring force provided by the baryon-photon fluid in the early Universe generates a set of wiggles in the spectrum $P(k)$ and an overall suppression due to diffusion. Both effects are discussed in detail by \citet{1998ApJ...496..605E} and \citet{1999ApJ...511....5E} who also provide a fitting formula for $T(k)$  in terms of the density parameters $\Omega_m$, $\Omega_b$, and the Hubble parameter $h$.

The spectrum $P(k)$ is normalized in such a way that it exhibits the variance $\sigma_8^2$ on the scale $R=8~\mathrm{Mpc}/h$,
\begin{equation}
\sigma^2_R = \int\frac{k^2\dd k}{2\pi^2}\:P(k) W^2(kR)
\end{equation}
with a Fourier transformed spherical top hat filter function, $W(x)=3j_1(x)/x$, where $j_\ell(x)$ is the spherical Bessel function of the first kind of order $\ell$ \citep{1972hmf..book.....A}.

\subsection{Structure growth}
The growth of density fluctuations in the cosmic matter distribution can be described as a self-gravitating hydrodynamical phenomenon, in the limit of Newtonian gravity. Homogeneous growth of the density field
\begin{equation}
\delta(\bmath{x},a)=D_+(a)\delta(\bmath{x},a=1)
\end{equation}
in the linear regime $\left|\delta\right|\ll 1$ is described by the growth function $D_+(a)$, which is the solution to the growth equation \citep{1997PhRvD..56.4439T, 1998ApJ...508..483W, 2003MNRAS.346..573L},
\begin{equation}
\frac{\dd^2}{\dd a^2}D_+(a) + \frac{1}{a}\left(3+\frac{\dd\ln H}{\dd\ln a}\right)\frac{\dd}{\dd a}D_+(a) = 
\frac{3}{2a^2}\Omega_m(a) D_+(a).
\label{eqn_growth}
\end{equation}

Nonlinear structure formation leads to a strongly enhanced structure growth on small scales, generates non-Gaussian features and, most importantly, wipes out BAO wiggles as features in the initial matter distribution. This can be understood in an intuitive way as corrections to the CDM spectrum in perturbation theory to order $n$ assume the shape of integrals over polyspectra up to order $2n$ \citep[which separate into a product of $n$ spectra by application of the Wick's theorem, see the review by][]{2002PhR...367....1B} and are therefore becoming insensitive to localized features that are not strongly influencing the normalization of $P(k)$ \citep{2005Natur.435..629S, 2006ApJ...651..619J, 2008JCAP...10..036P, 2008MPLA...23...25M, 2008PhRvD..77b3533C, 2009PASJ...61..321N, 2009ApJ...691..569J, 2012arXiv1204.6524J, 2012arXiv1205.2235A}. Since nonlinear structure formation affects small scales first, we will target BAO wiggles with 3dWL beginning at the largest wavelength before proceeding to successively shorter wavelengths.

\subsection{Weak gravitational lensing}
The weak lensing convergence $\kappa$ provides a weighted line-of-sight average of the matter density $\delta$ \citep[for reviews, see][]{2001PhR...340..291B, Munshi200867, 2008ARNPS..58...99H, 2010CQGra..27w3001B},
\begin{equation}
\kappa = \int_0^{\chi_H}\dd\chi\: W_\kappa(\chi)\delta,
\end{equation}
with the weak lensing efficiency $W_\kappa(\chi)$ as the weighting function,
\begin{equation}
W_\kappa(\chi) = \frac{3\Omega_m}{2\chi_H^2}\frac{D_+}{a}G(\chi)\chi,
\end{equation}
and the lensing efficiency weighted galaxy redshift distribution, rewritten in terms of comoving distance,
\begin{equation}
G(\chi) = 
\int_\chi^{\chi_H}\dd\chi^\prime\:n(\chi^\prime)\left(1-\frac{\chi^\prime}{\chi}\right).
\label{eqn_lensing_efficiency}
\end{equation}
$n(z)$ denotes a common parametrization of the redshift distribution of the lensed background galaxy sample,
\begin{equation}
n(z) = n_0\left(\frac{z}{z_0}\right)^2\exp\left(-\left(\frac{z}{z_0}\right)^\beta\right)\dd z
\quad\mathrm{with}\quad \frac{1}{n_0}=\frac{z_0}{\beta}\Gamma\left(\frac{3}{\beta}\right),
\label{eqn_redshift_distribution}
\end{equation}
which can be rewritten in terms of a distribution in comoving distance with the relation $n(\chi)\dd\chi = n(z)\dd z$ using $\dd\chi/\dd z = c/H(a)$. These expressions allow to carry out a Limber projection \citep{1954ApJ...119..655L} of the weak lensing convergence, which yields the angular convergence spectrum $C_\kappa(\ell)$, 
\begin{equation}
C_\kappa(\ell) = \int_0^{\chi_H}\frac{\dd\chi}{\chi^2}\:W_\kappa^2(\chi) P(k=\ell/\chi).
\label{eqn_ckappa}
\end{equation}
We will formulate our derivations in terms of the lensing convergence $\kappa$ instead of the observable shear $\gamma$ because it is a scalar quantity and possesses identical statistical properties. Eqn.~(\ref{eqn_ckappa}) illustrates why line-of-sight averaged weak lensing spectra are ineffective in picking up BAO wiggles (and is almost a repetition of the previous argument why nonlinear structure formation destroys BAO features): They provide only an integrated measure of the CDM spectrum $P(k)$ weighted with wide weighting functions $W_\kappa(\chi)$ that is very insensitive to local features of the spectrum such as BAO wiggles. This argument holds even for  advanced tomographic surveys \citep{1999ApJ...522L..21H,2004MNRAS.348..897T} and motivates the need of a 3-dimensional mapping of the cosmic matter distribution. With reference to \citet{2009MNRAS.399.1663G} and \citet{2010ApJ...719.1032K}, we would like to emphasize that weak lensing, due to its sensitivity to gravitational shear components perpendicular to the line of sight, will provide measurements of BAO wiggles in the transverse direction.

\section{3d weak lensing}\label{sect_3dwl}
The method of 3dWL was  introduced by \citet{2003MNRAS.343.1327H}, who proposed to include distances of lensed galaxies estimated from their photometric redshifts to infer the 3-dimensional {\em unprojected} tidal shear, i.e. the second derivatives of the gravitational potential perpendicular to the line-of-sight from distortions in the galaxies' ellipticity. Therefore, this approach differs from estimations of the angular line-of-sight averaged spectrum $C_\kappa(\ell)$ or corresponding tomographic spectra $C^{ij}_\kappa(\ell)$ in the important respect that the statistics of the full 3-dimensional matter distribution is inferred without any averaging of shears with the line-of-sight galaxy distribution, which has been performed in eqn.~(\ref{eqn_lensing_efficiency}). As such, 3dWL is particularly suited for the problem at hand, namely to provide a precise estimate of the amplitude of the dark matter power spectrum at the BAO wavelengths. Additionally, \citet{2003MNRAS.343.1327H} showed that if 3dWL is used for constraining $P(k)$ at a fixed cosmology, the smallest errors are expected in the BAO regime of the CDM spectrum.

In this section, we recapitulate the main results of 3dWL in terms of the weak lensing convergence in the Fourier-convention we prefer to work with; please also refer to \citet{2005PhRvD..72b3516C}, \citet{2007ApJS..172..239M}, \citet{2006MNRAS.373..105H}, \citet{2008PhRvD..77j3008K} for a detailed description of the theory, to \citet{2011MNRAS.411.2161M} for higher-order statistics through 3dWL and to \citet{2012MNRAS.422.3056A} for details of our numerical implementation. We assume spatial flatness and lensing in linearly evolving structures, which can be, in principle, relaxed from the 3dWL point of view \citep{2013arXiv1301.3673P}. The impact of systematic errors is nicely investigated by \citet{2008MNRAS.389..173K}, and for an application to observational data we refer the reader to \citet{2007MNRAS.376..771K}.

The most natural choice for carrying out a Fourier transform in spherical coordinates is a combination of spherical harmonics for the angular and spherical Bessel functions for the radial dependence. We can therefore write the transformation for the convergence $\kappa$ as
\begin{equation}
\kappa_{\ell m}(k)
\equiv 
\sqrt{\frac{2}{\pi}} \int\chi^2\dd\chi\: \dd\Omega\: \kappa(\chi,\btheta)\: j_\ell(k\chi) Y_{\ell m}^{*}(\btheta),
\end{equation}
\citep[see][]{1995MNRAS.276L..59B, 1995MNRAS.275..483H}, where $j_\ell$ and $Y_{\ell m}$ are, respectively, a spherical Bessel function of the first kind and a spherical harmonic, and  $\btheta \equiv (\theta, \varphi)$. There exist algorithms for fast computation of $\kappa_{\ell m}(k)$ \citep{2004MNRAS.353.1201P, 2012A&A...540A.115R, 2012A&A...540A..92L, 2012A&A...540A..60L}. Such a transformation is particularly convenient as the combination of $j_\ell$ and $Y_{\ell m}$ is an eigenfunction of the Laplacian in spherical coordinates, leading to a quite simple relationship between the coefficients of the density field $\delta_{\ell m}(k)$ and the lensing convergence $\kappa_{\ell m}(k)$ as the observable:
\begin{equation}
\kappa_{\ell m}(k) = 
\frac{3\Omega_m}{2\chi_H^2} 
\frac{\ell(\ell+1)}{2}
\frac{\eta_\ell(k,k^\prime)}{(k^\prime)^2}\delta_{\ell m}(k^\prime),
\end{equation}
with the lensing-induced mode coupling $\eta_\ell(k,k^\prime)$
\begin{equation}
\eta_\ell(k,k^\prime) = \frac{4}{\pi} 
\int_0^{\infty}{\chi^\prime}^2\dd\chi^\prime 
\int_0^{\chi^\prime}\dd\chi \:\frac{\chi^\prime - \chi}{\chi\chi^\prime}\frac{D_+}{a}\:j_\ell(k\chi')j_\ell(k^\prime\chi),
\end{equation}
with implicit assumption of the Einstein summation convention
\begin{equation}
X(k,k') \, Y(k',k'') \equiv \int_0^{\infty} {k^\prime}^2\dd k^\prime\: X(k,k') \, Y(k',k'').
\end{equation}

It is then possible to construct an estimator for $\kappa_{\ell m}(k)$ by including the uncertainty of the galaxy distance estimates coming from errors in the measurements of redshift. If we denote by $\chi$ the true radial coordinate of a galaxy, and by $\chi'$ the one inferred by its observed redshift $z^\prime=z(\chi^\prime)$, then they will be related by the probability $p(\chi'|\chi)$, which we assume to be Gaussian for simplicity:
\begin{equation}
p(\chi^\prime|\chi)\dd\chi = 
\frac{1}{\sqrt{2\pi\sigma_z^2}} \exp \left [ - \frac{(z(\chi) - z(\chi^\prime))^2}{2\sigma_z^2} \right ] \dd z^\prime, 
\end{equation}
where $\sigma_z$ is the width of the distribution and is assumed to be constant throughout the entire galaxy sample. Furthermore, galaxies receive a statistical weight according to their distribution in distance $n(\chi)\dd\chi$. Following the derivation in \cite{2003MNRAS.343.1327H}, we define the two additional matrices
\begin{eqnarray}
Z_\ell(k,k^\prime) &=& \frac{2}{\pi} \int{\chi^\prime}^2\dd\chi^\prime\:\int\dd\chi\:p(\chi^\prime|\chi)\:j_{\ell}(k^\prime\chi)j_\ell(k\chi'), \\
M_\ell(k,k^\prime) &=& \frac{2}{\pi} \int\chi^2\dd\chi\: n(\chi)\:j_\ell(k\chi) j_\ell(k^\prime\chi),
\end{eqnarray}
where $n(\chi)$ is the number density of galaxies, as defined in eqn.~(\ref{eqn_redshift_distribution}). These matrices describe the correlations in spherical Fourier modes generated by the measurement process: While $\eta_\ell(k,k^\prime)$ describes mode couplings due to weak lensing, $Z_\ell(k,k^\prime)$ and $M_\ell(k,k^\prime)$ define, respectively, the contributions in the mode couplings coming from redshift errors and from the galaxy distribution along the radial coordinate $\chi$. 

We restrict ourselves to observations of the entire sky. In this case, the expression for the estimator $\bar{\kappa}_{\ell m}$ of the convergence is then expected to be
\begin{equation}
\bar{\kappa}_{\ell m} (k) = \frac{3\Omega_m}{2\chi_H^2} \, \frac{\ell(\ell+1)}{2} \,\frac{B_\ell(k,k'')}{(k'')^2}\, \delta_{\ell m}(k'').
\label{eqn_kappa_estimator}
\end{equation}
where the mode-coupling matrix $B_\ell(k,k')$ describes two integrations over $k_1$ and $k_2$:
\begin{equation}
B_\ell(k,k'') = Z_\ell(k,k_1)M_\ell(k_1,k_2)\eta_\ell(k_2,k'').
\end{equation}

Since the average values of a field like $\kappa_{\ell m}(k)$ are zero for all-sky surveys, we can only infer information about any parameter the field may depend on by means of its covariance,
\begin{equation}
\label{eq_def_cov}
\langle \bar{\kappa}_{\ell m}(k) \, \bar{\kappa}_{\ell m}^*(k') \rangle =
 S_{\kappa,\ell}(k,k') + N_{\kappa,\ell}(k,k') \equiv C_{\kappa,\ell}(k,k')
\end{equation}
which consists of a signal term $S_{\kappa,\ell}(k,k')$ and a noise term $N_{\kappa,\ell}(k,k')$. The signal term $S_{\kappa,\ell}$ can be calculated directly from eqn.~(\ref{eqn_kappa_estimator}): 
\begin{equation}
\label{eq_covariance_signal}
S_{\kappa,\ell}(k,k') = 
\left(\frac{3\Omega_m}{2\chi_H^2}\right)^2 \left[ \frac{\ell(\ell+1)}{2} \right]^2 \frac{B_\ell(k,k'')}{(k^{''})^2}\frac{B_\ell(k',k'')}{(k^{''})^2}\:P_{\delta}(k''),
\end{equation}
with the abbreviations
\begin{eqnarray}
B_\ell(k,k'') & = & Z_\ell(k,k_1)\, M_\ell(k_1,k_2)\, \eta_\ell(k_2,k'') \\
B_\ell(k',k'') & = & Z_\ell(k',k_3)\, M_\ell(k_3,k_4)\, \eta_\ell(k_4,k'')
\end{eqnarray}
with implicit integration over $k_1$, $k_2$ and $k_3$, $k_4$. The corresponding noise part $N_{\kappa,\ell}$ is given by  
\begin{equation}
N_{\kappa,\ell}(k,k') = \frac{\sigma_{\epsilon}^2}{4} M_\ell(k,k'),
\end{equation}
which is proportional to the shape noise $\sigma_{\epsilon}^2$, namely the variance of the galaxy ellipticity distribution. It is important to notice that $N_{\kappa,\ell}$ is independent of cosmology or variations in the CDM spectrum $P(k)$. Intrinsic ellipticity correlations were neglected, which would greatly complicate the 3dWL description.

\section{Detecting BAO wiggles}\label{sect_wiggles}

\subsection{Construction of the Fisher matrix}
We choose a Fisher matrix approach  to determine how precisely  3dWL can constrain baryon acoustic oscillations in the matter power spectrum $P(k)$. The Fisher matrix is a square matrix whose elements are defined as the expectation values of the second derivative of the logarithmic likelihood with respect to the fiducial parameters $\theta_{\alpha}$ and $\theta_{\beta}$:
\begin{equation}
F_{\alpha \beta} = - \left \langle  \frac{\partial^2 \ln \likelihood}{\partial \theta_{\alpha} \partial \theta_{\beta}} \right \rangle.
\end{equation}
As a general statement, if the likelihood $\likelihood$ can be expressed as an $N$-dimensional Gaussian
\begin{equation}
\likelihood = \frac{1}{\sqrt{(2\pi)^N \det ( C)}} \exp \left( - \frac{1}{2} \, {\vec{x}}^{\:T}  C^{-1} \vec{x} \right),
\end{equation}
where $\vec x$ is a generic data vector and $C$ is the corresponding covariance, 
we can then write
\begin{equation} 
\label{eq_fisher_def}
F_{\alpha \beta} = \frac{1}{2} \trace \left [ (C^{-1}\; \partial_{\alpha}C) \times ( C^{-1} \; \partial_{\beta}C) \right ],
\end{equation}
or, equivalently,
\begin{equation}
F_{\alpha \beta} = \frac{1}{2} \trace \left [ \partial_{\alpha} \ln C \times \partial_{\beta} \ln C \right ],
\end{equation}
where $\partial_{\alpha}$ and $\partial_{\beta}$ stand for the derivatives with respect to the parameters $\theta_{\alpha}$ and $\theta_{\beta}$. Given a particular experimental framework, the Fisher matrix specifies what are the best errors to expect for the inferred parameters $\theta_{\alpha}$ via the Cram{\'e}r-Rao relation.

It can  be proved that, since $\ell$-measurements are independent in the case of full-sky coverage,
we can reformulate eqn.~(\ref{eq_fisher_def}):  we consider our estimator to be $\bar{\kappa}_{\ell m}(k)$ and its covariance as defined in eqn.~(\ref{eq_def_cov}), and find
\begin{equation} 
\label{eqn_fisher_summation}
F_{\alpha \beta} =\frac{f_\mathrm{sky}}{2} \sum_{\ell = \ell_{\mathrm{min}}}^{\ell_{\mathrm{max}}} (2 \ell + 1)\, \trace \left [ ({C_{\kappa,\ell}}^{-1} \; \partial_{\alpha}C_{\kappa,\ell}) \times ({C_{\kappa,\ell}}^{-1} \; \partial_{\beta}C_{\kappa,\ell}) \right ].
\end{equation}
\begin{figure} 
\resizebox{\hsize}{!}{\includegraphics{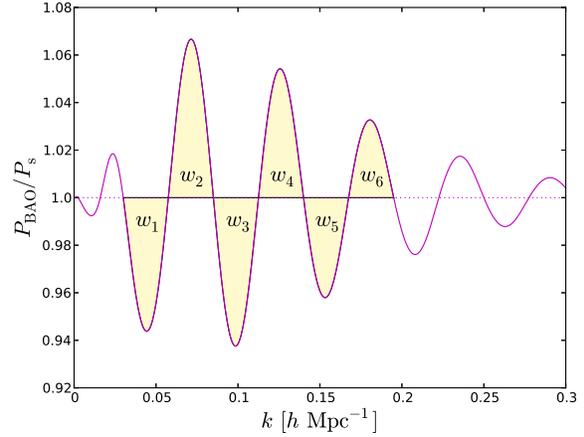}}
\caption{Ratio between the power spectrum with baryon acoustic oscillations $P(k)$ and the smooth power spectrum $P_{\mathrm{s}}(k)$. The largest wiggles in amplitude have been highlighted and labeled as $w_{\alpha}, \alpha = 1,...,6$. These are the wiggles used to parametrize the power spectrum in our Fisher matrix approach.}
\label{fig_wiggle_numbers}
\end{figure}
In our specific case, we consider the power spectrum as parametrized not by usual cosmological parameters, such as $\Omega_m$ or $\sigma_8$, but rather by its own wiggles amplitudes, namely by the values assumed by $P(k)$ in a range of $k$  where the wiggles $w_{\alpha}$ are located. In Fig.~\ref{fig_wiggle_numbers} we plot the wiggle-only power spectrum, i.e. the ratio between the power spectrum with baryon acoustic oscillations $P(k)$ and an equivalent, smoothed out spectrum that has the same shape as $P(k)$ but shows no oscillating feature, $P_{\mathrm{s}}(k)$. We highlight the wiggles that have been used to parametrize $P(k)$, $w_{\alpha}, \alpha = 1,...,6$.
By calculating the derivative $\partial_{\alpha}C$ of the covariance with respect to a variation of the amplitude of a maximum number of wiggles $n_{\mathrm{w}}$, we can build up (eqn.~\ref{eq_fisher_def}) a Fisher matrix $F_{\alpha \beta}$, where $\alpha, \beta = 1,...,n_{\mathrm{w}}$. Such a matrix carries information about the best errors to expect on the detection of each wiggle $w_{\alpha}$, $\alpha= 1,...,n_{\mathrm{w}}$, and the cross-correlations between inferred wiggle amplitudes. 
Given our aim, what we are actually performing in the calculation of $\partial_{\alpha}C$ is a functional derivative, also known as Fr{\'e}chet derivative. In fact, we can imagine the power spectrum  as depending on features, i.e. the wiggles in Fig.~\ref{fig_wiggle_numbers}. Each one of them can be approximated to a $\sin$-like function defined in a range of $k$ as wide  as $\lambda/2$, where $\lambda$ is the wavelength of the function itself. 
The covariance derivative is numerically estimated for one wiggle at a time as a finite difference:
\begin{equation}
\label{eq_cov_derivative}
\partial_{\alpha} C_{\kappa,\ell} =  \frac{C_{\kappa,\ell,\alpha}^+ - C_{\kappa,\ell,\alpha}^-}{2  \epsilon},
\end{equation}
where $C_{\kappa,\ell,\alpha}^{\pm}$ are the covariance matrices calculated using the power spectra $P^{\pm}_{\alpha}(k)$ and $\epsilon$ is an arbitrarily small number. The spectra  $P^{\pm}_{\alpha}(k)$ are equivalent to the original $P(k)$ for all $k$ of the domain, exception made for the wave numbers belonging to the interval $I_\alpha$ that corresponds to wiggle $w_{\alpha}$.
In this interval $P^{\pm}_{\alpha}(k)$ is then
\begin{equation}
\label{eq_p_variation}
P^{\pm}_{\alpha}(k) = P(k) \pm \epsilon  P(k).
\end{equation}
We would like to point out that, since what we are actually performing by means of the spectrum variation in eqn.~(\ref{eq_p_variation}) is in a way a logarithmic derivative of $C_\ell$, the denominator in eqn.~(\ref{eq_cov_derivative}) lacks a factor $P(k)$ and is therefore just two times the fraction of the spectrum used in the variation.
\begin{figure} 
\resizebox{\hsize}{!}{\includegraphics{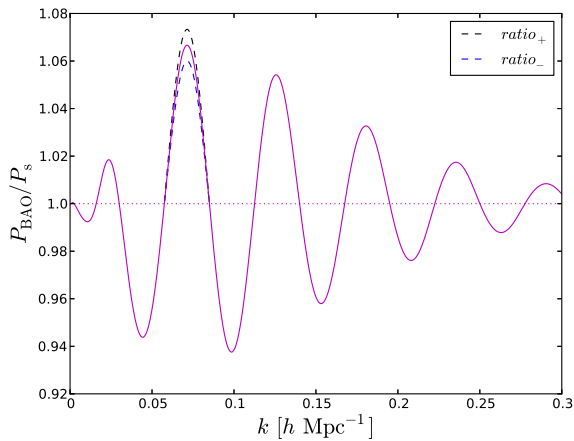}}
\caption{The ratio between  $P(k)$ and $P_{\mathrm{s}}(k)$. The figure shows a pictorial representation of the variation of the amplitude of one wiggle when calculating the derivative of the covariance matrix, $\partial_{\alpha = 2}C_{\ell}$.}
\label{fig_increment_nolabels}
\end{figure}
In Fig.~\ref{fig_increment_nolabels} we show a representation of an example of the variation performed in the calculation of the derivative of the covariance matrix, $\partial_{\alpha} C$; in this case the second wiggle, $w_2$, has been considered.

It is worth noticing that the Fisher-matrix approach for inferring the  error $\sigma_{\alpha}$ on the dark matter spectrum $P(k_\alpha)$ (where $k_{\alpha}$ are simply the $k \in I_{\alpha}$, for brevity) as a Gaussian standard deviation is perfectly justified because of the linearity of the lensing observable and the linearity of the random field, so we do not need to use Monte-Carlo sampling for evaluating the likelihood $\likelihood(P(k_\alpha))$ and to measure its widths $\sigma_{\alpha}$ from Monte-Carlo samples of the likelihood.

\begin{table}
\begin{center}
\begin{tabular}{llllll}
\hline\hline
			& $z_\mathrm{med}$ & $\bar{n}$ &$f_\mathrm{sky}$ & $\sigma_z$ & $\sigma_\epsilon$\\
\hline
EUCLID	& 0.9 & 30 & 0.5  & 0.1 & 0.3 \\
DES	& 0.7 & 10 & 0.1 & 0.02 & 0.3 \\
DEEP	& 1.5 & 40 & 0.1  & 0.05 & 0.3 \\
\hline
\end{tabular}
\end{center}
\caption{Basic survey characteristics used for the Fisher-analysis: median redshift $z_\mathrm{med}$ of the galaxy sample, galaxy density per squared arcminute $\bar{n}$,  sky coverage fraction $f_\mathrm{sky}$, redshift error $\sigma_z$ and shape measurement error $\sigma_\epsilon$ of the surveys EUCLID, DES and a hypothetical deep-reaching survey labeled DEEP.}
\label{table}
\end{table}

As noise sources for the inference of $P(k_\alpha)$, we consider a Gaussian shape measurement error $\sigma_\epsilon$ for the galaxy ellipticities, which are assumed to be intrinsically uncorrelated, and a Gaussian error $\sigma_z$ for the redshift determination uncertainty. Likewise, we work in the approximation of neglecting all geodesic effects \citep{1994A&A...287..349S,1994CQGra..11.2345S} like deviations form the Born approximation, lens-lens couplings \citep{2006JCAP...03..007S,2010A&A...523A..28K}, source clustering \citep{2002A&A...389..729S}, source-lens correlations \citep{2002MNRAS.330..365H}, and deviations from Newtonian gravity \citep{2004PhRvD..70b3515A}. While performing the necessary variations for computing the Fisher matrix, we keep all other cosmological parameters fixed and calculate everything using an $\ell$-range between $\ell_{\mathrm{min}} = 2$ and $\ell_{\mathrm{max}} = 100$ (please see the next section for a justification of this choice).  As surveys, we  consider the cases of EUCLID, DES, and a hypothetical deep-reaching galaxy survey we refer to by the name DEEP. The respective survey properties are summarized in Table~\ref{table}.
All the details concerning the numerical implementation used in the calculation of the covariance matrices can be found in \citet{2012MNRAS.422.3056A}.

\subsection{Statistical errors}

The error $\sigma_{\alpha}$ for inferring the amplitude of the CDM spectrum $P(k_\alpha)$ at wiggle positions $k_\alpha$ is given by the Cram{\'e}r-Rao relation,
\begin{equation}
\label{eq_cramer_rao}
\sigma_{\alpha}^2 = (F^{-1})_{\alpha\alpha},
\end{equation}
and
\begin{equation}
\sigma_{\alpha}^2 = 1/F_{\alpha\alpha},
\end{equation}
for marginalized and conditional likelihoods, respectively.

\begin{figure} 
\resizebox{\hsize}{!}{\includegraphics{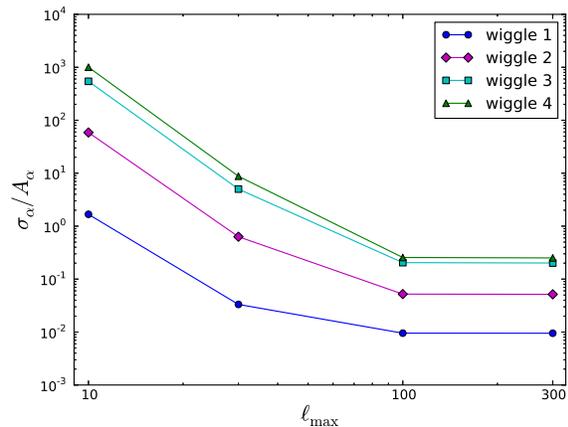}}
\caption{The marginalized errors on the first four wiggles, normalized with respect to their amplitude, as the maximum $\ell$ used for the calculation of the Fisher matrix increases. It shows that choosing $\ell_{\mathrm{max}} > 100$ brings no particular advantage in the precision of the wiggles measurement. In the calculation of the errors we considered the power spectrum as parametrized by the first 4 wiggles.} \label{fig_relerr_vs_lmax}
\end{figure}
Before carrying out our analysis for surveys like EUCLID, DES, and DEEP, we implement some tests in order to determine the optimal value for the maximum number of modes to be used in the calculation of the Fisher matrix, $\ell_{\mathrm{max}}$. Besides, we tried to find out how and by how much are the errors  sensitive  to some of the usual survey parameters, such as 
\begin{enumerate}
\item{the shape noise $\sigma_{\epsilon}$;}
\item{ the error $\sigma_{z}$ in the measurement of redshift;}
\item{the median redshift $z_{\mathrm{med}}$;}
\item{the fraction of sky coverage $f_{\mathrm{sky}}$.}
\end{enumerate}
Throughout these tests, when not stated otherwise, we make use of a default set of survey parameters such that $\sigma_z = 0.02$, $\sigma_{\epsilon} = 0.3$, $z_{\mathrm{med}} = 0.9$, $f_{\mathrm{sky}} = 0.4$ and  $\bar{n} = 20$. Additionally, we assumed we want to constrain simultaneously the first four wiggles (see Fig.~\ref{fig_wiggle_numbers}).

We start our investigation by determining the errors $\sigma_\alpha$, $\alpha = 1,2,3,4$, as the maximum number of modes $\ell_{\mathrm{max}}$ in the summation in eqn.~(\ref{eqn_fisher_summation}) increases. Please refer to Fig.~\ref{fig_relerr_vs_lmax} for a plot of the behavior of $\sigma_\alpha$ normalized to the oscillation amplitude $A_{\alpha}$, where $A_{\alpha}$ is defined as the maximum value of $|P(k) - P_{\mathrm{s}}(k)|$ for each wiggle. In particular, we considered $\ell_{\mathrm{max}} = 10,30,100,300$, and observe that, after $\ell_{\mathrm{max}} = 100$, there is practically no gain in the precision with which the first 4 wiggles would be constrained in a 3dWL approach. Therefore, we decide to stick to a maximum number of modes of $100$ for all the subsequent calculations. This is a fair approximation also from a theoretical point of view: in fact, extending too much the $\ell$ interval for the $F_{\alpha \beta}$ summation could make us fall out of the linear regime; in addition, the assumption of a Gaussian shape for the likelihood $\likelihood$ could not to be anymore reasonable in such a multipole range \citep{2003MNRAS.343.1327H}. 
\begin{figure} 
\resizebox{\hsize}{!}{\includegraphics{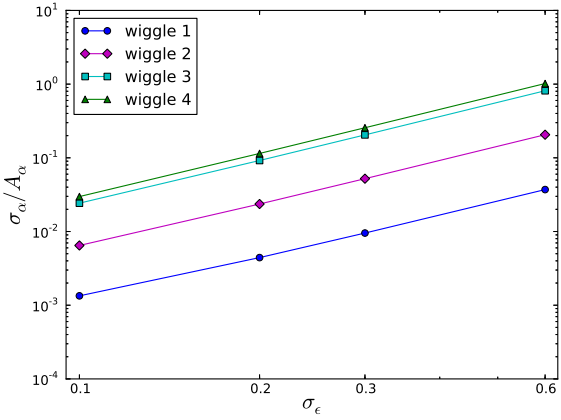}}
\caption{The relative marginalized error for the first 4 wiggles, as the shape noise $\sigma_{\epsilon}$ varies. Again, we assumed $\sigma_z = 0.02$, $\sigma_{\epsilon} = 0.3$, $z_{\mathrm{med}} = 0.9$, $f_{sky} = 0.4$ and $\bar{n} = 20$. As we expected, larger values of $\sigma_{\epsilon}$ bring along larger errors on the detection of the wiggles.}
\label{fig_sigmae_sens}
\end{figure}

In Fig.~\ref{fig_sigmae_sens}, we show what happens as soon as we keep all the survey parameters fixed and vary the shape noise $\sigma_{\epsilon}$. As one can expect, larger values of $\sigma_{\epsilon}$ lead to larger $\sigma_{\alpha}$, therefore to a greater uncertainty in the detection of the wiggles. Moreover, the rate at which $\sigma_{\alpha}$ grows with the shape noise seems to be of the type of a power law, and seems independent of the wiggle considered, at least for the wiggles sample we evaluated.
\begin{figure} 
\resizebox{\hsize}{!}{\includegraphics{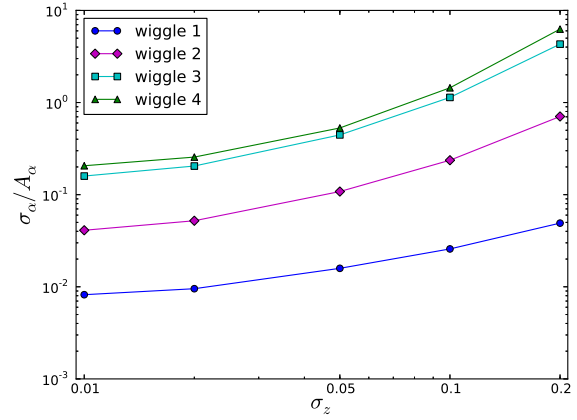}}
\caption{The relative marginalized error for the first 4 wiggles as a function of the error in the measurement of the photometric redshift, $\sigma_z$. The sensitivity of the errors on the value of $\sigma_z$ is not as steep as it is for the shape noise $\sigma_{\epsilon}$. In addition, the increasing rate of $\sigma_{\alpha}$ seems to change as we consider higher order wiggles.  }
\label{fig_sigmaz_sens}
\end{figure}
The situation turns out to be similar when the error in the determination of redshift $\sigma_{z}$ is considered, in Fig.~\ref{fig_sigmaz_sens}: increasing $\sigma_{z}$ still produces larger errors on all the wiggles under investigation, although here the relation is somewhat slower, especially as long as $\sigma_{z}\lesssim 0.1$; the relation also appears to be slightly dependent on the wiggle, becoming steeper as higher order oscillations are taken. In fact, by incrementing the error on redshift from $0.01$ to $0.1$, we get an error larger only by  a factor of $\sim 2$ on the first wiggle and by a factor of $\sim 8 $ on the second.
\begin{figure} 
\resizebox{\hsize}{!}{\includegraphics{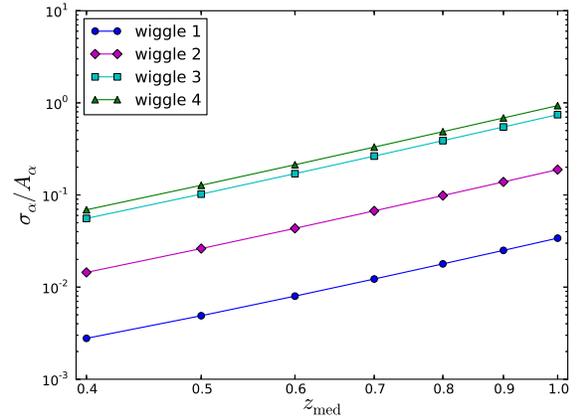}}
\caption{Relative marginalized errors on the first 4 wiggles when the median redshift $z_{\mathrm{med}}$ of the survey increases. Since all other parameters are kept fixed, especially the mean galaxy density per squared arcminute, $\bar n$, increasing $z_{\mathrm{med}}$ is equivalent to having more and more diluted surveys, where the same amount of galaxies is distributed along a deeper survey cone. }
\label{fig_zmed_sens}
\end{figure}
%
%
%
Additionally, the correspondence between $\sigma_{\alpha}$ and the median redshift of the survey (Fig.~\ref{fig_zmed_sens}) seems again like a power law that gives larger errors for an increasing $z_{\mathrm{med}}$, and is independent of the wiggle. Qualitatively, this trend makes sense in light of the fact that, as we increase the median redshift, we keep fixed all other survey parameters such as, for example, the galaxy density per squared arcminute $\bar n$. By doing so, we consider surveys where a number $\bar n$ of galaxies is distributed over a  deeper cone, meaning that we are actually sampling the 3D convergence field in a more diluted way, and therefore inheriting a larger noise. Naturally, varying the sky covarage propagates to the errors $\sigma_\alpha\propto 1/\sqrt{f_\mathrm{sky}}$.

\subsection{Detectability of BAO wiggles}
In this section, we would like to present the results obtained when estimating the best errors to expect on BAO wiggles  for the surveys EUCLID, DES, and DEEP (please see Table~\ref{table} for specifications).
\begin{figure} 
\resizebox{\hsize}{!}{\includegraphics{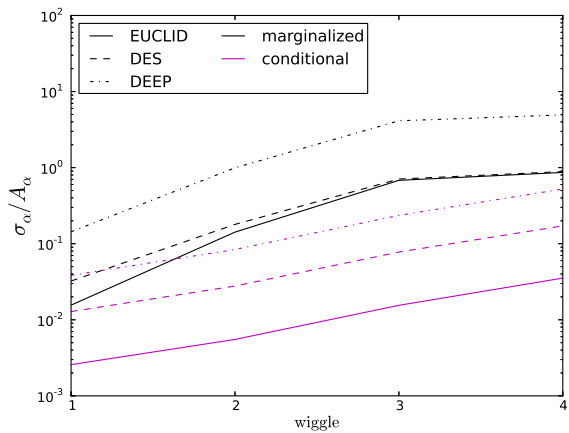}}
\caption{Conditional (magenta lines) and marginalized  relative errors (black lines) for the three surveys under investigations EUCLID (solid line), DES (dashed line) and DEEP (dash-dot line) as a function of the wiggles, when the first 4 oscillations are simultaneously constrained.}
\label{fig_conditional_marginalized}
\end{figure}
\begin{figure} 
\resizebox{\hsize}{!}{\includegraphics{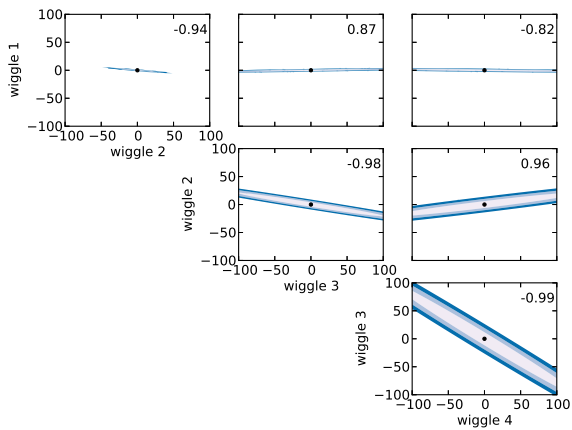}}
\caption{Confidence ellipses for the first four wiggles in a EUCLID-like survey, showing that the wiggles are indeed highly correlated. In fact, taking, for example, two consecutive wiggles, such as $w_2$ and $w_3$,  we see that by increasing the amplitude of $P(k)$ at the position of wiggle $2$, we must then have the amplitude at $w_3$ decreased in order to remain in the confidence region. The $x$- and $y$-axes show the variation of the wiggle in terms of percentage of its amplitude $A_\alpha$, the three contours areas correspond to $1-2-3\sigma$, and every panel shows the correlation coefficient in the upper-right corner.}
\label{fig_ellipses_EUCLID}
\end{figure}
\begin{figure} 
\resizebox{\hsize}{!}{\includegraphics{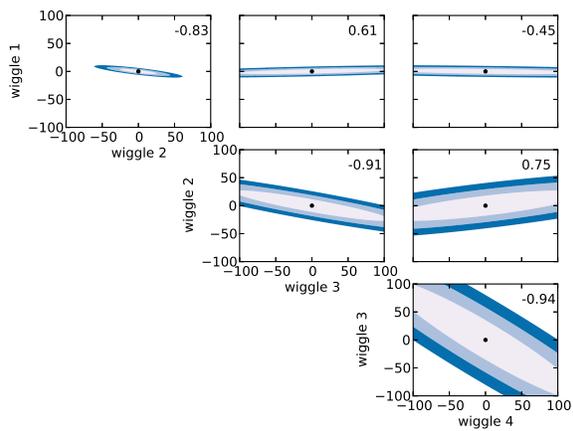}}
\caption{Fisher confidence ellipses for DES, when one tries to simultaneously constrain the first four wiggles. Again, contours areas correspond to $1-2-3\sigma$ and the number in every panel is the correlation coefficient; the axes represent variation of wiggles in terms of their amplitude fraction. Also in this case we can observe correlation between the amplitudes of $P(k)$ at different wiggles positions.}
\label{fig_ellipses_DES}
\end{figure}
\begin{figure} 
\resizebox{\hsize}{!}{\includegraphics{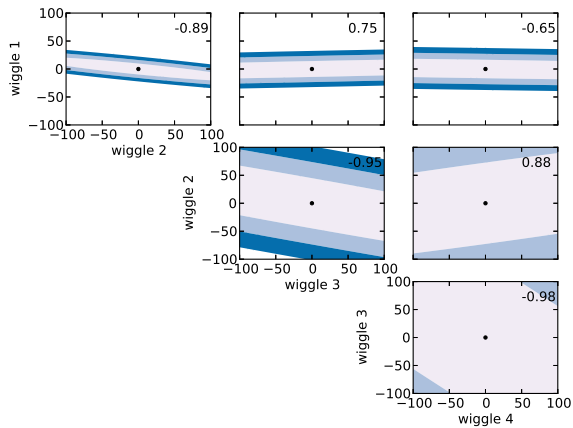}}
\caption{Confidence ellipses for the DEEP survey, with the variation of the oscillations  in terms of the wiggle amplitude. Again, we assumed we wanted to constrain jointly the first four wiggles,  contour areas stand for  $1-2-3\sigma$ and the correlation coefficient can be read in the panels.}
\label{fig_ellipses_DEEP}
\end{figure}

We started our analysis by calculating both marginalized ($\sigma_{\alpha} = \sqrt{(F^{-1})_{\alpha \alpha}}$) and conditional errors ($\sigma_{\alpha} = 1/\sqrt{F_{\alpha \alpha}}$) relative to the wiggle amplitude. These errors were computed  for the three types of surveys,  considering the first four oscillations, as shown in Fig.~\ref{fig_conditional_marginalized}. As we could expect, marginalized errors are always larger than the correspondent conditional ones, namely the  $\sigma_\alpha$ on each wiggle when we assume to know precisely all the other wiggle amplitudes. 

We continue the investigation considering the confidence ellipses calculated from the corresponding Fisher matrices obtained for the three surveys. We assume we are aiming to jointly constrain the first four wiggles and plot the results in Figs.~\ref{fig_ellipses_EUCLID}, \ref{fig_ellipses_DES}, and \ref{fig_ellipses_DEEP} for, respectively, EUCLID, DES, and DEEP. The sizes of the ellipses, whose contours stand for $1-2-3\sigma$, already tell us that, among the ones evaluated, EUCLID will probably be the survey with largest constraining power on the BAO wiggles. It is of particular interest noticing the orientation of the ellipses, or their correlation coefficients (upper-right corner in every panel), that tell us something about the interdepence between  different wiggles: in fact, neighboring wiggles are anti-correlated , i.e. increasing the amplitude of the power spectrum in correspondence to one oscillation would cause the $P(k)$ at the position of the adjacent wiggle to take smaller values in order to stay among the confidence region, and vice versa, whereas the opposite holds for alternated wiggles. 
\begin{figure} 
\resizebox{\hsize}{!}{\includegraphics{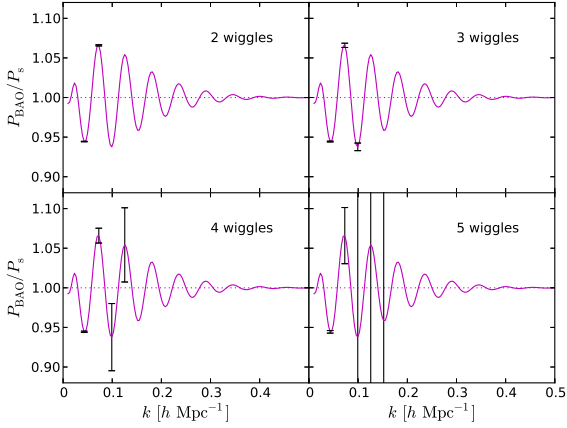}}
\caption{The bars show the marginalized errors  in the detection of the wiggles, normalized with respect to $P_{\mathrm{s}}(k)$, when the first 2 (upper left panel), 3 (upper right panel), 4 (bottom left panel), 5 (bottom right panel) wiggles are used to parametrize  the power spectrum. Here, we considered a EUCLID-like survey with $\sigma_z = 0.1$, $\sigma_{\epsilon} = 0.3$, $z_{\mathrm{med}} = 0.9$, $f_{\mathrm{sky}} = 0.5$ and $\bar{n} = 30$.}
\label{fig_multi_err_EUCLID}
\end{figure}
\begin{figure} 
\resizebox{\hsize}{!}{\includegraphics{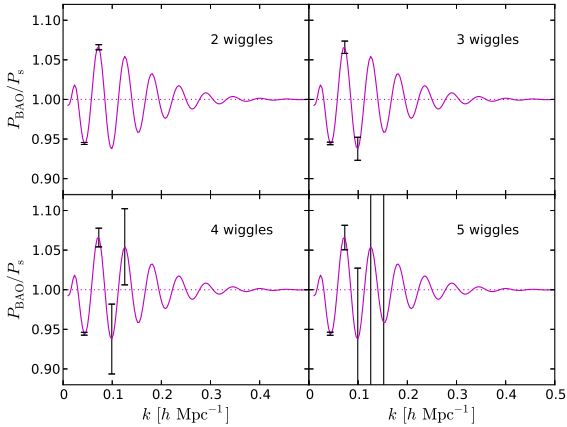}}
\caption{Marginalized errors on the detection of  BAO wiggles when one tries to detect the first 2 (upper left panel), 3 (upper right panel), 4 (bottom left panel), 5 (bottom right panel) wiggles at the same time, for a DES-like survey ($\sigma_z = 0.02$, $\sigma_{\epsilon} = 0.3$, $z_{\mathrm{med}} = 1.5$, $f_{\mathrm{sky}} = 0.12$ and $\bar{n} = 10$). Again, we considered the relative marginalized errors, dividing by $P_{\mathrm{s}}(k)$.}
\label{fig_multi_err_DES}
\end{figure}
\begin{figure} 
\resizebox{\hsize}{!}{\includegraphics{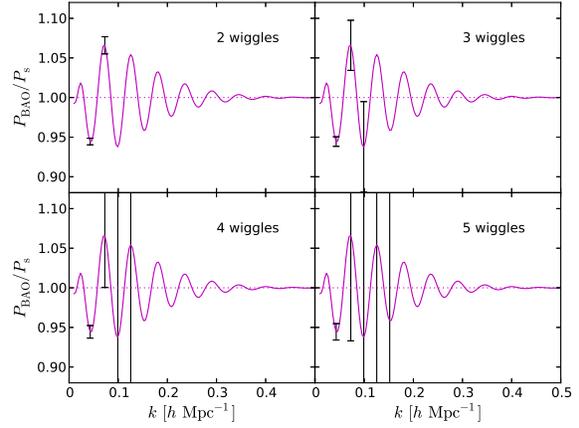}}
\caption{The marginalized errors on the detection of the wiggles in the power spectrum, relative to $P_{\mathrm{s}}(k)$, for the hypothetical survey DEEP, characterized by the parameters $\sigma_z = 0.02$, $\sigma_{\epsilon} = 0.3$, $z_{\mathrm{med}} = 1.5$, $f_{\mathrm{sky}} = 0.1$ and $\bar{n} = 40$.}
\label{fig_multi_err_DEEP}
\end{figure}

In order to better understand whether the constraining power of the three surveys will allow us to detect any oscillatory feature in the CDM power spectrum, we plot the $\sigma_\alpha$  obtained from  the Cram{\'e}r-Rao relation in eqn.~(\ref{eq_cramer_rao}) as error bars in the usual wiggle-only power spectrum for DEEP (Fig.~\ref{fig_multi_err_DEEP}), DES (Fig.~\ref{fig_multi_err_DES}), and EUCLID (Fig.~\ref{fig_multi_err_EUCLID}). Since what is shown is a ratio between $P(k)$ and a smooth spectrum,  the $\sigma_{\alpha}$ have of course also been normalized with respect to $P_{\mathrm{s}}(k)$. The four different panels show how the errors change when we try to jointly constrain the first 2, 3, 4, or 5 wiggles with a 3dWL approach. 

What these and the following plots show, first of all, is an expected feature: as we increment the number of wiggles we expect to simultaneously examine, the precision with which the amplitudes $P(k_\alpha)$ would be measured gets poorer and poorer for all the oscillations. Our purpose would then be to evaluate how many BAO wiggles one is allowed to constrain before the errors on them become too large.
It can be seen that all three surveys would allow for quite good constraints on the first 2 wiggles. The hypothetical survey DEEP already shows error bars of the order of the wiggle amplitude $A_{\alpha}$ when the first 3 wiggles are considered, and the errors become much larger than $A_{\alpha}$ ($\alpha > 1$) as soon as one tries to detect 4 or more wiggles (Fig.~\ref{fig_multi_err_DEEP}). On the other hand, DES and EUCLID give a better performance, allowing for up to 4 wiggles to be simultaneously constrained, with EUCLID giving smaller errors overall (Fig.~\ref{fig_multi_err_EUCLID}).

\begin{figure} 
\resizebox{\hsize}{!}{\includegraphics{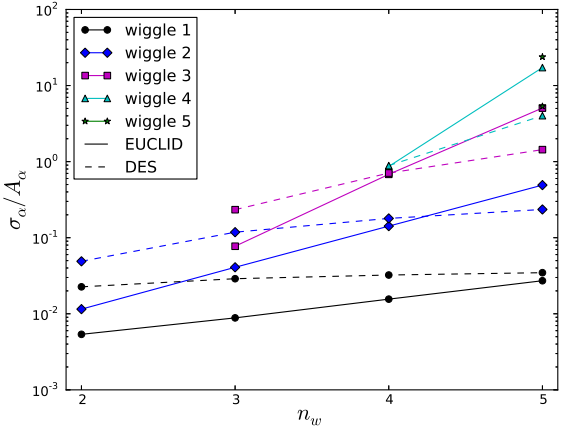}}
\caption{Relative marginalized errors (with respect to the wiggles amplitude) as a function of the maximum number of wiggles that we try to simultaneously detect, $n_w$. Here, we show the results for a EUCLID-like survey (solid lines) and a DES-like survey (dashed lines). Different colors and point types correspond to the different wiggles on which the error $\sigma_{\alpha}$ is calculated. The relative errors coming from  a DES-like survey turn out to be larger than EUCLID's as long as we try to observe a maximum of four wiggles. DES gives a better performance for $n_w >4$, but the $\sigma_{\alpha}$ on the measurement of the higher order wiggles prove to be too large with respect to their amplitude $A_{\alpha}$, anyway.}
\label{fig_wiggle_num_DES_vs_EUCLID_new}
\end{figure}

\begin{figure} 
\resizebox{\hsize}{!}{\includegraphics{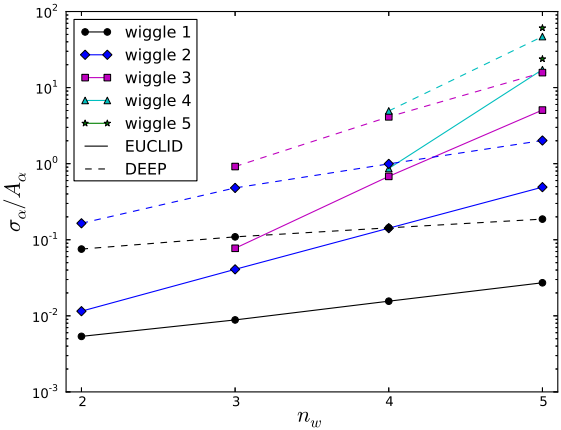}}
\caption{Relative marginalized errors as a function of $n_w$ for a EUCLID-like survey (solid lines) and a DEEP-like survey (dashed lines). The relative $\sigma_{\alpha}$ for a DEEP survey appear to be always larger than the ones obtained from EUCLID, independently of the maximum number of wiggles $n_w$ we want to constrain.}
\label{fig_wiggle_num_DEEP_vs_EUCLID}
\end{figure}

A better comparison between the three surveys can be carried out analyzing Figs.~\ref{fig_wiggle_num_DEEP_vs_EUCLID} and \ref{fig_wiggle_num_DES_vs_EUCLID_new}, where we plotted relative errors $\sigma_{\alpha} / A_{\alpha}$ as functions of the maximum number of wiggles we want to jointly constrain, $n_{\mathrm{w}}$, and we collate results coming from, respectively, DEEP and EUCLID, and DES, and EUCLID. 
It becomes straightforward that a DEEP-like survey cannot compete against EUCLID:  the relative errors coming from DEEP are always larger than the latter's, independently of the maximum number of wiggles to be constrained, and remain safely under the unity only for the first 2 wiggles.

The situation is better for DES, which gives results quite similar to EUCLID's up to $n_{\mathrm{w}} = 4$, and an even higher performance afterwards, since its $\sigma_{\alpha}$ are increasing at a smaller pace with respect to EUCLID's. This is not very useful, anyway, as the relative errors become larger than 1 for $n_{\mathrm{w}}>4$ for both surveys. 

Concluding, EUCLID seems to grant the best results, allowing for the simultaneous detection of up to 4 wiggles with expected errors that are smaller than the ones predicted for both DES and, of course, DEEP.
%

\section{Summary and conclusions}\label{sect_summary}
Subject of this paper has been a statistical investigation on whether future weak lensing surveys are able to detect baryon acoustic oscillations in the cosmic matter distribution by application of the 3d weak lensing method. For a fixed $w$CDM cosmology, we have estimated the statistical precision $\sigma_{\alpha}$ on the amplitude of the CDM spectrum $P(k)$ at the BAO wiggle positions in a Fisher-matrix approach. Throughout, we worked under the assumption of Gaussian statistics, independent Fourier modes and in the limit of weak lensing. Noise sources were idealized and consisted in independent Gaussian-distributed shape-noise measurements for the lensed background galaxy sample, as well as a Gaussian error for the redshift determination. As surveys, we considered the cases of EUCLID, DES and a hypothetical deep-reaching survey DEEP.

\begin{enumerate}
\item{We have constructed the Fisher matrix considering our model as parametrized by the amplitudes of the CDM power spectrum at the baryon acoustic oscillations anticipated positions. In particular, we started taking the two BAO wiggles with largest amplitude and progressively increased the number of oscillations considered. Keeping the cosmology fixed to a standard $w$CDM parameter choice, we carried out variations of $P(k)$ that preserved its wiggle-shape in those wave number intervals; we then estimated the Fisher matrix accordingly, in order to quantify whether the statistical power of future weak lensing surveys suffices to place bounds on the amplitudes of the considered harmonics. By means of the Cram{\'e}r-Rao relation, we calculated the best errors $\sigma_\alpha$ to expect for the amplitudes of $P(k)$ at wiggle positions. }
\item{The sensitivity of $\sigma_\alpha$ with respect to some typical survey-parameters was tested. In particular, we considered  the shape noise $\sigma_\epsilon$, the redshift error $\sigma_z$, the median redshift $z_{\mathrm{med}}$, and the sky coverage $f_{\mathrm{sky}}$. We found that, as expected, increasing the uncertainty in the estimate of either the redshift or the galaxy shapes brings a larger error in the inference of the presence of wiggles, and that the sensitivity of   these errors on $\sigma_z$ is  less pronounced for small values of $\sigma_z$, although it grows as soon as we consider higher order wiggles or large $\sigma_z$. 
An increase of $z_{\mathrm{med}}$ leads as well to larger errors on the wiggles amplitudes, as one would expect from considering less and less populated surveys; for the same reason, a wider sky coverage, i.e. larger $f_{\mathrm{sky}}$, yields to higher precision in constraining the wiggle amplitudes $\sigma_\alpha\propto 1/\sqrt{f_\mathrm{sky}}$.  \\
Overall, we may conclude that the volume of a survey seem to be overcoming the importance of a high precision in the redshift measurement of galaxies, at least for $\sigma_{z}<0.1-0.2$.}
\item{Finally, we evaluated the $\sigma_\alpha$ for the surveys under investigation and found that, among them, EUCLID gave the best results, potentially allowing for the detection of up to the first four BAO wiggles with a good statistical confidence. Given our tests on the sensitivity of the errors on $P(k_\alpha)$ to certain survey parameters, we may conclude that EUCLID's good performance is probably due to the volume of the survey in terms of total galaxy number and sky coverage, that seem to prevail over the negative effects brought by the  error on redshift measurements, $\sigma_z$, quite larger than the ones predicted for the other two surveys.}
\end{enumerate}

Given these results, we conclude that measurements of BAO wiggles based on future weak lensing data are entirely possible, and avoid issues related to galaxy biasing and redshift-space distortions. We forecast a detection of the first four wiggles with EUCLID and DES by applying 3dWL techniques. Future developments from our side include estimates of the precision that can be reached on inferring dark energy density and equation of state by including the estimate of the BAO scale at low redshifts probed by lensing to the estimates at intermediate redshift provided by galaxy surveys and those at high redshifts such as the CMB. Additionally, we are investigating the impact of systematical errors on the estimation process from 3dWL-data and biases in the estimation of BAO-wiggle amplitudes.

\section*{Acknowledgements}
Our work was supported by the German Research Foundation (DFG) within the framework of the excellence initiative through the Heidelberg Graduate School of Fundamental Physics. AG receives support from the Graduate School of Fundamental Physics (GSFP$+$), in addition AG would like to acknowledge support from the International Max-Planck Research School for Astronomy and Cosmic Physics. We would like to thank first of all Youness Ayaita and Maik Weber for the precious contribution to this work. We are also grateful to Matthias Bartelmann,  Angelos Kalovidouris, Federica Capranico, and Philipp M. Merkel for their advice and suggestions, and of course to Alan Heavens, who was so kind to answer our questions at the Transregio Winter School in Passo del Tonale.

\bibliography{bibtex/aamnem,bibtex/references}

\begin{thebibliography}{}

\bibitem[\protect\citeauthoryear{{Abramowitz} \& {Stegun}}{{Abramowitz} \&
  {Stegun}}{1972}]{1972hmf..book.....A}
{Abramowitz} M.,  {Stegun} I.~A.,  1972, {Handbook of Mathematical Functions}.
Handbook of Mathematical Functions, New York: Dover, 1972

\bibitem[\protect\citeauthoryear{{Acquaviva}, {Baccigalupi} \&
  {Perrotta}}{{Acquaviva} et~al.}{2004}]{2004PhRvD..70b3515A}
{Acquaviva} V.,  {Baccigalupi} C.,    {Perrotta} F.,  2004, \prd, 70, 023515

\bibitem[\protect\citeauthoryear{{Angulo}, {Baugh}, {Frenk} \&
  {Lacey}}{{Angulo} et~al.}{2008}]{2008MNRAS.383..755A}
{Angulo} R.~E.,  {Baugh} C.~M.,  {Frenk} C.~S.,    {Lacey} C.~G.,  2008,
  \mnras, 383, 755

\bibitem[\protect\citeauthoryear{{Anselmi} \& {Pietroni}}{{Anselmi} \&
  {Pietroni}}{2012}]{2012arXiv1205.2235A}
{Anselmi} S.,  {Pietroni} M.,  2012, ArXiv e-prints 1205.2235

\bibitem[\protect\citeauthoryear{{Ayaita}, {Sch{\"a}fer} \& {Weber}}{{Ayaita}
  et~al.}{2012}]{2012MNRAS.422.3056A}
{Ayaita} Y.,  {Sch{\"a}fer} B.~M.,    {Weber} M.,  2012, \mnras, 422, 3056

\bibitem[\protect\citeauthoryear{{Ballinger}, {Heavens} \&
  {Taylor}}{{Ballinger} et~al.}{1995}]{1995MNRAS.276L..59B}
{Ballinger} W.~E.,  {Heavens} A.~F.,    {Taylor} A.~N.,  1995, \mnras, 276, L59

\bibitem[\protect\citeauthoryear{{Bartelmann}}{{Bartelmann}}{2010}]{2010CQGra..27w3001B}
{Bartelmann} M.,  2010, Classical and Quantum Gravity, 27, 233001

\bibitem[\protect\citeauthoryear{{Bartelmann} \& {Schneider}}{{Bartelmann} \&
  {Schneider}}{2001}]{2001PhR...340..291B}
{Bartelmann} M.,  {Schneider} P.,  2001, \physrep, 340, 291

\bibitem[\protect\citeauthoryear{{Bassett} \& {Hlozek}}{{Bassett} \&
  {Hlozek}}{2010}]{2010deot.book..246B}
{Bassett} B.,  {Hlozek} R.,  2010, {Baryon acoustic oscillations}.
p.~246

\bibitem[\protect\citeauthoryear{{Bennett}, {Kogut}, {Hinshaw}, {Banday},
  {Wright}, {Gorski}, {Wilkinson}, {Weiss} \& {et}}{{Bennett}
  et~al.}{1994}]{1994ApJ...436..423B}
{Bennett} C.~L.,  {Kogut} A.,  {Hinshaw} G.,  {Banday} A.~J.,  {Wright} E.~L.,
  {Gorski} K.~M.,  {Wilkinson} D.~T.,  {Weiss} R.,    {et} a.,  1994, \apj,
  436, 423

\bibitem[\protect\citeauthoryear{{Bernardeau}, {Colombi}, {Gazta{\~n}aga} \&
  {Scoccimarro}}{{Bernardeau} et~al.}{2002}]{2002PhR...367....1B}
{Bernardeau} F.,  {Colombi} S.,  {Gazta{\~n}aga} E.,    {Scoccimarro} R.,
  2002, \physrep, 367, 1

\bibitem[\protect\citeauthoryear{{Beutler}, {Blake}, {Colless}, {Jones},
  {Staveley-Smith}, {Campbell}, {Parker}, {Saunders} \& {Watson}}{{Beutler}
  et~al.}{2011}]{2011MNRAS.416.3017B}
{Beutler} F.,  {Blake} C.,  {Colless} M.,  {Jones} D.~H.,  {Staveley-Smith} L.,
   {Campbell} L.,  {Parker} Q.,  {Saunders} W.,    {Watson} F.,  2011, \mnras,
  416, 3017

\bibitem[\protect\citeauthoryear{{Busca}, {Delubac}, {Rich}, {Bailey},
  {Font-Ribera}, {Kirkby}, {Le Goff}, {Pieri} \& {et}}{{Busca}
  et~al.}{2012}]{2012arXiv1211.2616B}
{Busca} N.~G.,  {Delubac} T.,  {Rich} J.,  {Bailey} S.,  {Font-Ribera} A.,
  {Kirkby} D.,  {Le Goff} J.-M.,  {Pieri} M.~M.,    {et} a.,  2012, ArXiv
  e-prints 1211.2616

\bibitem[\protect\citeauthoryear{{Cabr{\'e}} \& {Gazta{\~n}aga}}{{Cabr{\'e}} \&
  {Gazta{\~n}aga}}{2011}]{2011MNRAS.412L..98C}
{Cabr{\'e}} A.,  {Gazta{\~n}aga} E.,  2011, \mnras, 412, L98

\bibitem[\protect\citeauthoryear{{Castro}, {Heavens} \& {Kitching}}{{Castro}
  et~al.}{2005}]{2005PhRvD..72b3516C}
{Castro} P.~G.,  {Heavens} A.~F.,    {Kitching} T.~D.,  2005, \prd, 72, 023516

\bibitem[\protect\citeauthoryear{{Crocce} \& {Scoccimarro}}{{Crocce} \&
  {Scoccimarro}}{2008}]{2008PhRvD..77b3533C}
{Crocce} M.,  {Scoccimarro} R.,  2008, \prd, 77, 023533

\bibitem[\protect\citeauthoryear{{Desjacques}, {Crocce}, {Scoccimarro} \&
  {Sheth}}{{Desjacques} et~al.}{2010}]{2010PhRvD..82j3529D}
{Desjacques} V.,  {Crocce} M.,  {Scoccimarro} R.,    {Sheth} R.~K.,  2010,
  \prd, 82, 103529

\bibitem[\protect\citeauthoryear{{Dolney}, {Jain} \& {Takada}}{{Dolney}
  et~al.}{2006}]{2006MNRAS.366..884D}
{Dolney} D.,  {Jain} B.,    {Takada} M.,  2006, \mnras, 366, 884

\bibitem[\protect\citeauthoryear{{Eisenstein} \& {Hu}}{{Eisenstein} \&
  {Hu}}{1998}]{1998ApJ...496..605E}
{Eisenstein} D.~J.,  {Hu} W.,  1998, \apj, 496, 605

\bibitem[\protect\citeauthoryear{{Eisenstein} \& {Hu}}{{Eisenstein} \&
  {Hu}}{1999}]{1999ApJ...511....5E}
{Eisenstein} D.~J.,  {Hu} W.,  1999, \apj, 511, 5

\bibitem[\protect\citeauthoryear{{Eisenstein}, {Seo}, {Sirko} \&
  {Spergel}}{{Eisenstein} et~al.}{2007}]{2007ApJ...664..675E}
{Eisenstein} D.~J.,  {Seo} H.-J.,  {Sirko} E.,    {Spergel} D.~N.,  2007, \apj,
  664, 675

\bibitem[\protect\citeauthoryear{{Eisenstein}, {Zehavi}, {Hogg}, {Scoccimarro},
  {Blanton}, {Nichol}, {Scranton}, {Seo} \& {et}}{{Eisenstein}
  et~al.}{2005}]{2005ApJ...633..560E}
{Eisenstein} D.~J.,  {Zehavi} I.,  {Hogg} D.~W.,  {Scoccimarro} R.,  {Blanton}
  M.~R.,  {Nichol} R.~C.,  {Scranton} R.,  {Seo} H.-J.,    {et} a.,  2005,
  \apj, 633, 560

\bibitem[\protect\citeauthoryear{{Gazta{\~n}aga}, {Cabr{\'e}}, {Castander},
  {Crocce} \& {Fosalba}}{{Gazta{\~n}aga} et~al.}{2009}]{2009MNRAS.399..801G}
{Gazta{\~n}aga} E.,  {Cabr{\'e}} A.,  {Castander} F.,  {Crocce} M.,
  {Fosalba} P.,  2009, \mnras, 399, 801

\bibitem[\protect\citeauthoryear{{Gazta{\~n}aga}, {Cabr{\'e}} \&
  {Hui}}{{Gazta{\~n}aga} et~al.}{2009}]{2009MNRAS.399.1663G}
{Gazta{\~n}aga} E.,  {Cabr{\'e}} A.,    {Hui} L.,  2009, \mnras, 399, 1663

\bibitem[\protect\citeauthoryear{{Hamana}, {Colombi}, {Thion}, {Devriendt},
  {Mellier} \& {Bernardeau}}{{Hamana} et~al.}{2002}]{2002MNRAS.330..365H}
{Hamana} T.,  {Colombi} S.~T.,  {Thion} A.,  {Devriendt} J.~E.~G.~T.,
  {Mellier} Y.,    {Bernardeau} F.,  2002, \mnras, 330, 365

\bibitem[\protect\citeauthoryear{{Heavens}}{{Heavens}}{2003}]{2003MNRAS.343.1327H}
{Heavens} A.,  2003, \mnras, 343, 1327

\bibitem[\protect\citeauthoryear{{Heavens}, {Kitching} \& {Taylor}}{{Heavens}
  et~al.}{2006}]{2006MNRAS.373..105H}
{Heavens} A.~F.,  {Kitching} T.~D.,    {Taylor} A.~N.,  2006, \mnras, 373, 105

\bibitem[\protect\citeauthoryear{{Heavens} \& {Taylor}}{{Heavens} \&
  {Taylor}}{1995}]{1995MNRAS.275..483H}
{Heavens} A.~F.,  {Taylor} A.~N.,  1995, \mnras, 275, 483

\bibitem[\protect\citeauthoryear{{Hinshaw}, {Nolta}, {Bennett}, {Bean},
  {Dor{\'e}}, {Greason}, {Halpern}, {Hill} \& {et}}{{Hinshaw}
  et~al.}{2007}]{2007ApJS..170..288H}
{Hinshaw} G.,  {Nolta} M.~R.,  {Bennett} C.~L.,  {Bean} R.,  {Dor{\'e}} O.,
  {Greason} M.~R.,  {Halpern} M.,  {Hill} R.~S.,    {et} a.,  2007, \apjs, 170,
  288

\bibitem[\protect\citeauthoryear{{Hinshaw}, {Spergel}, {Verde}, {Hill},
  {Meyer}, {Barnes}, {Bennett}, {Halpern} \& {et}}{{Hinshaw}
  et~al.}{2003}]{2003ApJS..148..135H}
{Hinshaw} G.,  {Spergel} D.~N.,  {Verde} L.,  {Hill} R.~S.,  {Meyer} S.~S.,
  {Barnes} C.,  {Bennett} C.~L.,  {Halpern} M.,    {et} a.,  2003, \apjs, 148,
  135

\bibitem[\protect\citeauthoryear{{Hoekstra} \& {Jain}}{{Hoekstra} \&
  {Jain}}{2008}]{2008ARNPS..58...99H}
{Hoekstra} H.,  {Jain} B.,  2008, Annual Review of Nuclear and Particle
  Science, 58, 99

\bibitem[\protect\citeauthoryear{{Hu}}{{Hu}}{1999}]{1999ApJ...522L..21H}
{Hu} W.,  1999, \apjl, 522, L21

\bibitem[\protect\citeauthoryear{{Hu} \& {Sugiyama}}{{Hu} \&
  {Sugiyama}}{1996}]{1996ApJ...471..542H}
{Hu} W.,  {Sugiyama} N.,  1996, \apj, 471, 542

\bibitem[\protect\citeauthoryear{{Jeong} \& {Komatsu}}{{Jeong} \&
  {Komatsu}}{2006}]{2006ApJ...651..619J}
{Jeong} D.,  {Komatsu} E.,  2006, \apj, 651, 619

\bibitem[\protect\citeauthoryear{{Jeong} \& {Komatsu}}{{Jeong} \&
  {Komatsu}}{2009}]{2009ApJ...691..569J}
{Jeong} D.,  {Komatsu} E.,  2009, \apj, 691, 569

\bibitem[\protect\citeauthoryear{{J{\"u}rgens} \& {Bartelmann}}{{J{\"u}rgens}
  \& {Bartelmann}}{2012}]{2012arXiv1204.6524J}
{J{\"u}rgens} G.,  {Bartelmann} M.,  2012, ArXiv e-prints 1204.6524

\bibitem[\protect\citeauthoryear{{Kazin}, {Blanton}, {Scoccimarro}, {McBride}
  \& {Berlind}}{{Kazin} et~al.}{2010}]{2010ApJ...719.1032K}
{Kazin} E.~A.,  {Blanton} M.~R.,  {Scoccimarro} R.,  {McBride} C.~K.,
  {Berlind} A.~A.,  2010, \apj, 719, 1032

\bibitem[\protect\citeauthoryear{{Kazin}, {Blanton}, {Scoccimarro}, {McBride},
  {Berlind}, {Bahcall}, {Brinkmann}, {Czarapata} \& {et}}{{Kazin}
  et~al.}{2010}]{2010ApJ...710.1444K}
{Kazin} E.~A.,  {Blanton} M.~R.,  {Scoccimarro} R.,  {McBride} C.~K.,
  {Berlind} A.~A.,  {Bahcall} N.~A.,  {Brinkmann} J.,  {Czarapata} P.,    {et}
  a.,  2010, \apj, 710, 1444

\bibitem[\protect\citeauthoryear{{Kitching}, {Heavens}, {Taylor}, {Brown},
  {Meisenheimer}, {Wolf}, {Gray} \& {Bacon}}{{Kitching}
  et~al.}{2007}]{2007MNRAS.376..771K}
{Kitching} T.~D.,  {Heavens} A.~F.,  {Taylor} A.~N.,  {Brown} M.~L.,
  {Meisenheimer} K.,  {Wolf} C.,  {Gray} M.~E.,    {Bacon} D.~J.,  2007,
  \mnras, 376, 771

\bibitem[\protect\citeauthoryear{{Kitching}, {Heavens}, {Verde}, {Serra} \&
  {Melchiorri}}{{Kitching} et~al.}{2008}]{2008PhRvD..77j3008K}
{Kitching} T.~D.,  {Heavens} A.~F.,  {Verde} L.,  {Serra} P.,    {Melchiorri}
  A.,  2008, \prd, 77, 103008

\bibitem[\protect\citeauthoryear{{Kitching}, {Taylor} \& {Heavens}}{{Kitching}
  et~al.}{2008}]{2008MNRAS.389..173K}
{Kitching} T.~D.,  {Taylor} A.~N.,    {Heavens} A.~F.,  2008, \mnras, 389, 173

\bibitem[\protect\citeauthoryear{{Krause} \& {Hirata}}{{Krause} \&
  {Hirata}}{2010}]{2010A&A...523A..28K}
{Krause} E.,  {Hirata} C.~M.,  2010, \aap, 523, A28

\bibitem[\protect\citeauthoryear{{Labatie}, {Starck} \&
  {Lachi{\`e}ze-Rey}}{{Labatie} et~al.}{2012}]{2012ApJ...746..172L}
{Labatie} A.,  {Starck} J.~L.,    {Lachi{\`e}ze-Rey} M.,  2012, \apj, 746, 172

\bibitem[\protect\citeauthoryear{{Lanusse}, {Rassat} \& {Starck}}{{Lanusse}
  et~al.}{2012}]{2012A&A...540A..92L}
{Lanusse} F.,  {Rassat} A.,    {Starck} J.-L.,  2012, \aap, 540, A92

\bibitem[\protect\citeauthoryear{{Larson}, {Dunkley}, {Hinshaw}, {Komatsu},
  {Nolta}, {Bennett}, {Gold}, {Halpern} \& {et}}{{Larson}
  et~al.}{2011}]{2011ApJS..192...16L}
{Larson} D.,  {Dunkley} J.,  {Hinshaw} G.,  {Komatsu} E.,  {Nolta} M.~R.,
  {Bennett} C.~L.,  {Gold} B.,  {Halpern} M.,    {et} a.,  2011, \apjs, 192, 16

\bibitem[\protect\citeauthoryear{{Leistedt}, {Rassat}, {R{\'e}fr{\'e}gier} \&
  {Starck}}{{Leistedt} et~al.}{2012}]{2012A&A...540A..60L}
{Leistedt} B.,  {Rassat} A.,  {R{\'e}fr{\'e}gier} A.,    {Starck} J.-L.,  2012,
  \aap, 540, A60

\bibitem[\protect\citeauthoryear{{Leonard}, {Dup{\'e}} \& {Starck}}{{Leonard}
  et~al.}{2012}]{2012A&A...539A..85L}
{Leonard} A.,  {Dup{\'e}} F.-X.,    {Starck} J.-L.,  2012, \aap, 539, A85

\bibitem[\protect\citeauthoryear{{Limber}}{{Limber}}{1954}]{1954ApJ...119..655L}
{Limber} D.~N.,  1954, \apj, 119, 655

\bibitem[\protect\citeauthoryear{{Linder} \& {Jenkins}}{{Linder} \&
  {Jenkins}}{2003}]{2003MNRAS.346..573L}
{Linder} E.~V.,  {Jenkins} A.,  2003, \mnras, 346, 573

\bibitem[\protect\citeauthoryear{{Massey}, {Rhodes}, {Leauthaud}, {Capak},
  {Ellis}, {Koekemoer}, {R{\'e}fr{\'e}gier}, {Scoville} \& {et}}{{Massey}
  et~al.}{2007}]{2007ApJS..172..239M}
{Massey} R.,  {Rhodes} J.,  {Leauthaud} A.,  {Capak} P.,  {Ellis} R.,
  {Koekemoer} A.,  {R{\'e}fr{\'e}gier} A.,  {Scoville} N.,    {et} a.,  2007,
  \apjs, 172, 239

\bibitem[\protect\citeauthoryear{{Matarrese} \& {Pietroni}}{{Matarrese} \&
  {Pietroni}}{2008}]{2008MPLA...23...25M}
{Matarrese} S.,  {Pietroni} M.,  2008, Modern Physics Letters A, 23, 25

\bibitem[\protect\citeauthoryear{{Mehta}, {Cuesta}, {Xu}, {Eisenstein} \&
  {Padmanabhan}}{{Mehta} et~al.}{2012}]{2012arXiv1202.0092M}
{Mehta} K.~T.,  {Cuesta} A.~J.,  {Xu} X.,  {Eisenstein} D.~J.,    {Padmanabhan}
  N.,  2012, ArXiv e-prints 1202.0092

\bibitem[\protect\citeauthoryear{{Meiksin}, {White} \& {Peacock}}{{Meiksin}
  et~al.}{1999}]{1999MNRAS.304..851M}
{Meiksin} A.,  {White} M.,    {Peacock} J.~A.,  1999, \mnras, 304, 851

\bibitem[\protect\citeauthoryear{{Montanari} \& {Durrer}}{{Montanari} \&
  {Durrer}}{2011}]{2011PhRvD..84b3522M}
{Montanari} F.,  {Durrer} R.,  2011, \prd, 84, 023522

\bibitem[\protect\citeauthoryear{{Munshi}, {Heavens} \& {Coles}}{{Munshi}
  et~al.}{2011}]{2011MNRAS.411.2161M}
{Munshi} D.,  {Heavens} A.,    {Coles} P.,  2011, \mnras, 411, 2161

\bibitem[\protect\citeauthoryear{Munshi, Valageas, van Waerbeke \&
  Heavens}{Munshi et~al.}{2008}]{Munshi200867}
Munshi D.,  Valageas P.,  van Waerbeke L.,    Heavens A.,  2008, Physics
  Reports, 462, 67

\bibitem[\protect\citeauthoryear{{Nishimichi}, {Ohmuro}, {Nakamichi}, {Taruya},
  {Yahata}, {Shirata}, {Saito}, {Nomura}, {Yamamoto} \& {Suto}}{{Nishimichi}
  et~al.}{2007}]{2007PASJ...59.1049N}
{Nishimichi} T.,  {Ohmuro} H.,  {Nakamichi} M.,  {Taruya} A.,  {Yahata} K.,
  {Shirata} A.,  {Saito} S.,  {Nomura} H.,  {Yamamoto} K.,    {Suto} Y.,  2007,
  \pasj, 59, 1049

\bibitem[\protect\citeauthoryear{{Nishimichi}, {Shirata}, {Taruya}, {Yahata},
  {Saito}, {Suto}, {Takahashi}, {Yoshida}, {Matsubara}, {Sugiyama}, {Kayo},
  {Jing} \& {Yoshikawa}}{{Nishimichi} et~al.}{2009}]{2009PASJ...61..321N}
{Nishimichi} T.,  {Shirata} A.,  {Taruya} A.,  {Yahata} K.,  {Saito} S.,
  {Suto} Y.,  {Takahashi} R.,  {Yoshida} N.,  {Matsubara} T.,  {Sugiyama} N.,
  {Kayo} I.,  {Jing} Y.,    {Yoshikawa} K.,  2009, \pasj, 61, 321

\bibitem[\protect\citeauthoryear{{Nolta}, {Dunkley}, {Hill}, {Hinshaw},
  {Komatsu}, {Larson}, {Page}, {Spergel} \& {et}}{{Nolta}
  et~al.}{2009}]{2009ApJS..180..296N}
{Nolta} M.~R.,  {Dunkley} J.,  {Hill} R.~S.,  {Hinshaw} G.,  {Komatsu} E.,
  {Larson} D.,  {Page} L.,  {Spergel} D.~N.,    {et} a.,  2009, \apjs, 180, 296

\bibitem[\protect\citeauthoryear{{Padmanabhan}, {Schlegel}, {Seljak},
  {Makarov}, {Bahcall}, {Blanton}, {Brinkmann}, {Eisenstein} \&
  {et}}{{Padmanabhan} et~al.}{2007}]{2007MNRAS.378..852P}
{Padmanabhan} N.,  {Schlegel} D.~J.,  {Seljak} U.,  {Makarov} A.,  {Bahcall}
  N.~A.,  {Blanton} M.~R.,  {Brinkmann} J.,  {Eisenstein} D.~J.,    {et} a.,
  2007, \mnras, 378, 852

\bibitem[\protect\citeauthoryear{{Padmanabhan}, {Xu}, {Eisenstein}, {Scalzo},
  {Cuesta}, {Mehta} \& {Kazin}}{{Padmanabhan}
  et~al.}{2012}]{2012arXiv1202.0090P}
{Padmanabhan} N.,  {Xu} X.,  {Eisenstein} D.~J.,  {Scalzo} R.,  {Cuesta} A.~J.,
   {Mehta} K.~T.,    {Kazin} E.,  2012, ArXiv e-prints 1202.0090

\bibitem[\protect\citeauthoryear{{Parejko}, {Sunayama}, {Padmanabhan}, {Wake},
  {Berlind}, {Bizyaev}, {Blanton}, {Bolton} \& {et}}{{Parejko}
  et~al.}{2012}]{2012arXiv1211.3976P}
{Parejko} J.~K.,  {Sunayama} T.,  {Padmanabhan} N.,  {Wake} D.~A.,  {Berlind}
  A.~A.,  {Bizyaev} D.,  {Blanton} M.,  {Bolton} A.~S.,    {et} a.,  2012,
  ArXiv e-prints 1211.3976

\bibitem[\protect\citeauthoryear{Parkinson, Riemer-S\o{}rensen, Blake, Poole,
  Davis, Brough, Colless, Contreras \& et al.}{Parkinson
  et~al.}{2012}]{wigglez}
Parkinson D.,  Riemer-S\o{}rensen S.,  Blake C.,  Poole G.~B.,  Davis T.~M.,
  Brough S.,  Colless M.,  Contreras C.,    et al. 2012, Phys. Rev. D, 86,
  103518

\bibitem[\protect\citeauthoryear{{Percival}, {Burkey}, {Heavens}, {Taylor},
  {Cole}, {Peacock}, {Baugh}, {Bland-Hawthorn} \& {et}}{{Percival}
  et~al.}{2004}]{2004MNRAS.353.1201P}
{Percival} W.~J.,  {Burkey} D.,  {Heavens} A.,  {Taylor} A.,  {Cole} S.,
  {Peacock} J.~A.,  {Baugh} C.~M.,  {Bland-Hawthorn} J.,    {et} a.,  2004,
  \mnras, 353, 1201

\bibitem[\protect\citeauthoryear{{Percival}, {Cole}, {Eisenstein}, {Nichol},
  {Peacock}, {Pope} \& {Szalay}}{{Percival} et~al.}{2007}]{2007MNRAS.381.1053P}
{Percival} W.~J.,  {Cole} S.,  {Eisenstein} D.~J.,  {Nichol} R.~C.,  {Peacock}
  J.~A.,  {Pope} A.~C.,    {Szalay} A.~S.,  2007, \mnras, 381, 1053

\bibitem[\protect\citeauthoryear{{Percival}, {Nichol}, {Eisenstein},
  {Weinberg}, {Fukugita}, {Pope}, {Schneider}, {Szalay} \& {et}}{{Percival}
  et~al.}{2007}]{2007ApJ...657...51P}
{Percival} W.~J.,  {Nichol} R.~C.,  {Eisenstein} D.~J.,  {Weinberg} D.~H.,
  {Fukugita} M.,  {Pope} A.~C.,  {Schneider} D.~P.,  {Szalay} A.~S.,    {et}
  a.,  2007, \apj, 657, 51

\bibitem[\protect\citeauthoryear{{Percival}, {Reid}, {Eisenstein}, {Bahcall},
  {Budavari}, {Frieman}, {Fukugita}, {Gunn} \& {et}}{{Percival}
  et~al.}{2010}]{2010MNRAS.401.2148P}
{Percival} W.~J.,  {Reid} B.~A.,  {Eisenstein} D.~J.,  {Bahcall} N.~A.,
  {Budavari} T.,  {Frieman} J.~A.,  {Fukugita} M.,  {Gunn} J.~E.,    {et} a.,
  2010, \mnras, 401, 2148

\bibitem[\protect\citeauthoryear{{Pietroni}}{{Pietroni}}{2008}]{2008JCAP...10..036P}
{Pietroni} M.,  2008, \jcap, 10, 36

\bibitem[\protect\citeauthoryear{{Pratten} \& {Munshi}}{{Pratten} \&
  {Munshi}}{2013}]{2013arXiv1301.3673P}
{Pratten} G.,  {Munshi} D.,  2013, ArXiv e-prints

\bibitem[\protect\citeauthoryear{{Rassat} \& {Refregier}}{{Rassat} \&
  {Refregier}}{2012}]{2012A&A...540A.115R}
{Rassat} A.,  {Refregier} A.,  2012, \aap, 540, A115

\bibitem[\protect\citeauthoryear{{Schneider}, {van Waerbeke} \&
  {Mellier}}{{Schneider} et~al.}{2002}]{2002A&A...389..729S}
{Schneider} P.,  {van Waerbeke} L.,    {Mellier} Y.,  2002, \aap, 389, 729

\bibitem[\protect\citeauthoryear{{Seitz} \& {Schneider}}{{Seitz} \&
  {Schneider}}{1994}]{1994A&A...287..349S}
{Seitz} S.,  {Schneider} P.,  1994, \aap, 287, 349

\bibitem[\protect\citeauthoryear{{Seitz}, {Schneider} \& {Ehlers}}{{Seitz}
  et~al.}{1994}]{1994CQGra..11.2345S}
{Seitz} S.,  {Schneider} P.,    {Ehlers} J.,  1994, Classical and Quantum
  Gravity, 11, 2345

\bibitem[\protect\citeauthoryear{{Seljak} \& {Zaldarriaga}}{{Seljak} \&
  {Zaldarriaga}}{1996}]{1996ApJ...469..437S}
{Seljak} U.,  {Zaldarriaga} M.,  1996, \apj, 469, 437

\bibitem[\protect\citeauthoryear{{Seo} \& {Eisenstein}}{{Seo} \&
  {Eisenstein}}{2003}]{2003ApJ...598..720S}
{Seo} H.-J.,  {Eisenstein} D.~J.,  2003, \apj, 598, 720

\bibitem[\protect\citeauthoryear{{Shapiro} \& {Cooray}}{{Shapiro} \&
  {Cooray}}{2006}]{2006JCAP...03..007S}
{Shapiro} C.,  {Cooray} A.,  2006, Journal of Cosmology and Astro-Particle
  Physics, 3, 7

\bibitem[\protect\citeauthoryear{{Springel}, {White}, {Jenkins}, {Frenk},
  {Yoshida}, {Gao}, {Navarro}, {Thacker} \& {et}}{{Springel}
  et~al.}{2005}]{2005Natur.435..629S}
{Springel} V.,  {White} S.~D.~M.,  {Jenkins} A.,  {Frenk} C.~S.,  {Yoshida} N.,
   {Gao} L.,  {Navarro} J.,  {Thacker} R.,    {et} a.,  2005, \nat, 435, 629

\bibitem[\protect\citeauthoryear{{Sutherland}}{{Sutherland}}{2012}]{2012MNRAS.426.1280S}
{Sutherland} W.,  2012, \mnras, 426, 1280

\bibitem[\protect\citeauthoryear{{Takada} \& {Jain}}{{Takada} \&
  {Jain}}{2004}]{2004MNRAS.348..897T}
{Takada} M.,  {Jain} B.,  2004, \mnras, 348, 897

\bibitem[\protect\citeauthoryear{{Taruya}, {Nishimichi}, {Saito} \&
  {Hiramatsu}}{{Taruya} et~al.}{2009}]{2009PhRvD..80l3503T}
{Taruya} A.,  {Nishimichi} T.,  {Saito} S.,    {Hiramatsu} T.,  2009, \prd, 80,
  123503

\bibitem[\protect\citeauthoryear{{Turner} \& {White}}{{Turner} \&
  {White}}{1997}]{1997PhRvD..56.4439T}
{Turner} M.~S.,  {White} M.,  1997, \prd, 56, 4439

\bibitem[\protect\citeauthoryear{{Wang} \& {Steinhardt}}{{Wang} \&
  {Steinhardt}}{1998}]{1998ApJ...508..483W}
{Wang} L.,  {Steinhardt} P.~J.,  1998, \apj, 508, 483

\bibitem[\protect\citeauthoryear{{Wright}, {Bennett}, {Gorski}, {Hinshaw} \&
  {Smoot}}{{Wright} et~al.}{1996}]{1996ApJ...464L..21W}
{Wright} E.~L.,  {Bennett} C.~L.,  {Gorski} K.,  {Hinshaw} G.,    {Smoot}
  G.~F.,  1996, \apjl, 464, L21

\bibitem[\protect\citeauthoryear{{Zhao}, {Saito}, {Percival}, {Ross},
  {Montesano}, {Viel}, {Schneider}, {Ernst} \& {et}}{{Zhao}
  et~al.}{2012}]{2012arXiv1211.3741Z}
{Zhao} G.-B.,  {Saito} S.,  {Percival} W.~J.,  {Ross} A.~J.,  {Montesano} F.,
  {Viel} M.,  {Schneider} D.~P.,  {Ernst} D.~J.,    {et} a.,  2012, ArXiv
  e-prints 1211.3741

\end{thebibliography}
\bibliographystyle{mn2e}

\appendix

\bsp

\label{lastpage}

\end{document}